# CL API: Real-Time Closed-Loop Interactions with Biological Neural Networks

**David Hogan**
Cortical Labs
Melbourne, Australia, 3000
david@corticallabs.com

**Andrew Doherty**
Cortical Labs
Melbourne, Australia, 3000
andrew@corticallabs.com

**Boon Kien Khoo**
Cortical Labs
Selangor, Malaysia, 47100
boonkien@corticallabs.com

**Johnson Zhou**
Cortical Labs
Melbourne, Australia, 3000
johnson@corticallabs.com

**Richard Salib**
Cortical Labs
Melbourne, Australia, 3000
richard@corticallabs.com

**James Stewart**
Cortical Labs
Melbourne, Australia, 3000
james@corticallabs.com

**Kiaran Lawson**
Cortical Labs
Melbourne, Australia, 3000
kiaran@corticallabs.com

**Alon Loeffler**
Cortical Labs
Melbourne, Australia, 3000
alon@corticallabs.com

**Brett J. Kagan**
Cortical Labs
Melbourne, Australia, 3000
brett@corticallabs.com

## Abstract

Biological neural networks (BNNs) are increasingly explored for their rich dynamics, parallelism, and adaptive behavior. Beyond understanding their function as a scientific endeavour, a key focus has been using these biological systems as a novel computing substrate. However, BNNs can only function as reliable information-processing systems if inputs are delivered in a temporally and structurally consistent manner. In practice, this requires stimulation with precisely controlled structure, microsecond-scale timing, multi-channel synchronization, and the ability to observe and respond to neural activity in real-time. Existing approaches to interacting with BNNs face a fundamental trade-off: they either depend on low-level hardware mechanisms, imposing prohibitive complexity for rapid iteration, or they sacrifice temporal and structural control, undermining consistency and reproducibility – particularly in closed-loop experiments. The Cortical Labs Application Programming Interface (CL API) enables real-time, sub-millisecond closed-loop interactions with BNNs. Taking a contract-based API design approach, the CL API provides users with precise stimulation semantics, transactional admission, deterministic ordering, and explicit synchronization guarantees. This contract is presented through a declarative Python interface, enabling non-expert programmers to express complex stimulation and closed-loop behavior without managing low-level scheduling or hardware details. Ultimately, the CL API provides an accessible and reproducible foundation for real-time experimentation with BNNs, supporting both fundamental biological research and emerging neurocomputing applications.

## 1 Introduction

With rapidly growing interest in alternative computing methods, the study of how biological neurons process information and function in a way as to produce intelligence holds unique promise. This promise has also resulted in attempts to use the biological substrate itself as biological neural network (BNN) that can be harnessed directly in controllable ways as exemplified in the fields of Synthetic Biological Intelligence (SBI) [1–3] and Organoid Intelligence [4–8]. Biological neurons are highly power-efficient and sample-efficient, requiring a fraction of a percentage of either resource required by artificial intelligence (AI) systems [9, 10]. Beyond their extreme power- and sample-efficiency [11], BNNs display rich parallel dynamics and the capacity for robust synaptic plasticity – and functional connectivity changes unavailable to silicon von Neumann architectures [12–21].

Interacting with BNN, including for neurocomputing, requires a rigorous digital-biological interface. Stimulation must be *temporally and structurally consistent*, building on foundational demonstrations of writing information into neural networks [22, 23]. Every input – including stim subcomponent count, duration, amplitude, polarity and timing – must be precisely enacted and outcomes faithfully recorded [24–26]. Without such time-correct control and the ability to define the stimulation into, and read the activity from, BNN, the software-tissue interface becomes unreliable and experimental outcomes fragile.

Advancing this technology requires three pillars:

1. **Wetware:** Ethical, scalable, and specific neural cells;
2. **Hardware:** Viable interfaces maintaining cell health;
3. **Software:** Real-time frameworks for closed-loop algorithms.

Substantial work using synthetic biology to differentiate pluripotent stem cells into functional neural cells has provided a pathway to resolve the first point [27–29]. The development of scalable hardware has also recently been reported that resolves the second point [30], although such areas remain an active area requiring further development. Yet the third challenge has remained unsolved. As such, this paper aims to describe a method to address the third and final requirement.

## 2 Current Approaches to Interacting with BNNs

While BNNs may be explored using optogenetic [31, 32] or chemical methods [33], electrophysiological methods have grown to be the primary method for deep insight into neural dynamics at both the single cell[34] and population level [35]. Fundamentally, biological neurons generate measurable electrical pulses during action potentials. Consequently, electrophysiological dynamics can be measured across a neural population using devices such as microelectrode arrays (MEAs) [24–26]. First demonstrations that *in vitro* BNNs are responsive and would adapt to electrical stimulation via MEA stimulation established that not only can activity be recorded, but external electrical stimulation can be used to write information into BNNs [22, 23]. The use of open-loop paradigms that examine how BNNs may transform information encoded via electrical information, often called reservoir computing (RC), has also been informative. Evidence supports that these open-loop patterns of structured stimulation will induce meaningful changes in neural activity, suggesting even relatively simple BNNs have capabilities to distinguish different patterns and even engage in blind-source separation tasks [36–38]. Finally, work exploring closed-loop algorithms has established that these algorithms can rapidly induce robust and complex synaptic plasticity and functional connectivity changes, often with highly nuanced population-wide dynamics [11, 18–21]. However, there is significant variability in the methods and reproducibility across these experiments. Moreover, key limitations exist ranging from long-latencies and high jitter, to inflexible setups limiting additional meaningful controls. Although hardware differences exist and are widely discussed, e.g., [6, 39–41], underlying software differences are less frequently described in this area. Yet without transparent, specific, and controllable software to facilitate electrophysiological interactions with BNNs, progress will be stymied.

Algorithms facilitating sub-millisecond interactions with BNNs have previously been demonstrated. However, these systems typically interact directly with a Field Programmable Gate Array (FPGA) and can have slow development times with limited transparency [42]. Approaches that exclusively use object-oriented programming languages such as Python allow rapid algorithmic iteration but are limited in stimulation generation and exhibit relatively slow and variable response latencies of >60 ms [43]. One promising approach is to combine these ideas into a unified framework. For example, have an accessible object-oriented programming language be interpreted via embedded Linux systems to an FPGA to provide the necessary control and sub-millisecond latencies required for real-time closed-loop systems, without increasing the burden on implementation. Other approaches require choosing between the speed of direct hardware (sub-millisecond performance) and the flexibility of high-level software environments, which often suffer from higher latency and limited transparency [44]. Moreover, while various systems have been developed with promise [8, 45–48], none provide a well-defined, high-level API that enables non-expert users to rapidly iterate while retaining precise control over stimulation semantics in an extensible, interoperable, and modular framework.

As such, a tension between *temporal correctness* and *usability* currently bottlenecks research. Although existing commercial systems claim to offer stability, they often lack open APIs or modifiability that can function without sacrificing temporal correctness or having longer than ideal latencies. Conversely, ad-hoc research setups suffer from a reproducibility crisis due to significant methodological variability.

## 3 A Contract-Based Solution

To resolve this bottleneck, we introduce the Cortical Labs Application Programming Interface (CL API), a Python interface providing a reliable abstraction for real-time BNN interaction. Its central contribution is not specific hardware, but a clearly defined *execution contract* governing timing, ordering, synchronization, and failure semantics. This contract treats neural interactions – including both stimulation and access to real-time analog to digital converter (ADC) data – as a time-correct, information-bearing operation, guaranteeing closed-loop correctness through transactional admission, deterministic ordering, and observable execution outcomes.

The CL API exposes these guarantees via a declarative programming model. Users specify experimental intent, including consistent stimulation patterns, synchronization points, and closed-loop logic, without managing low-level scheduling or input-output (I/O). By separating semantic intent from execution details, the API enables accessible experimentation for interdisciplinary researchers while preserving the temporal correctness required for research.

We describe the CL1 platform as a reference implementation of this contract. The CL1 tightly integrates a user-space library, a Linux kernel driver, and an FPGA to ensure that high-level Python CL API calls can be executed with negligible jitter and latency measured in microseconds. A compatible simulator further supports development without physical hardware, enabling reproducible experimentation across simulation and deployment.

This paper makes three primary contributions:

1. A formal programming contract for real-time, closed-loop interaction with biological neural networks, specifying timing, ordering, synchronization, and failure semantics independent of implementation.
2. A declarative programming model that exposes this contract to users, enabling precise specification of stimulation and recording behavior without requiring low-level scheduling or hardware control.
3. An empirical demonstration of feasibility, via a reference implementation showing that the proposed semantics can be satisfied in practice under real-time and hardware resource constraints.

By formalizing the semantics of temporally and structurally consistent stimulation, the CL API aims to raise the standards of neurocomputing software infrastructure and provide a foundation for reproducible, scalable experimentation.

## 4 Design Principles

When designing the CL API, we established two complementary design pillars: *ease of use*, prioritizing the researcher's workflow through flexibility and abstraction, and *precision*, strictly enforcing the temporal guarantees required for valid biological interaction.

### 4.1 Ease of Use

**Put the user first.** Functionality without usability acts as a barrier to scientific adoption. This reality drove the decision to implement the CL API in Python, the standard language for data science and machine learning. By leveraging the Python ecosystem, the API allows researchers to integrate seamlessly with existing tools for data processing, debugging, and visualization [49]. We prioritize intuitive function signatures and a declarative style, ensuring that the complexity of the underlying execution mechanisms is abstracted away. The goal is to minimize the cognitive load required to operate the system, allowing researchers to focus on experimental logic rather than implementation details.

**Modularity.** Interacting with *in vitro* BNNs involves navigating significant uncertainty; far more is unknown than known regarding how different cell types process information. Given this biological variability, a rigid software tool would limit discovery. The CL API embraces modularity, providing basic computational building blocks that can be assembled to meet unforeseen experimental requirements. This extensibility allows users to rapidly iterate, identifying key biological dynamics in one experiment and immediately incorporating those insights into the next, without requiring changes to the underlying platform.

**Transparency.** A disconnect often exists between the engineers who build neuro-interface systems and the biologists who use them. An MEA system may be architected by hardware specialists, but the end-user is often a cell biologist who may lack the specific domain knowledge to audit the hardware's performance. Furthermore, existing systems often obscure the precise timing of command delivery or raw data streams. The CL API prioritizes transparency by ensuring that the execution semantics of the system are observable. Providing full visibility into how and when commands are executed is critical for deep research, where distinguishing between biological artifacts and system behaviors is essential.

## 4.2 Precision

**Temporal Fidelity.** Biological neurons utilize precise timing codes to process information. Mechanisms such as Spike-Timing-Dependent Plasticity (STDP) rely on correlations between pre- and post-synaptic events on the order of milliseconds. Consequently, precise temporal control must be a fundamental architectural constraint. The CL API enforces a strict execution contract (formalized in Section 5) in which stimulation waveforms are generated with microsecond-scale determinism. This ensures that the stimulation structure (duration, amplitude, and polarity) remains consistent, preventing software jitter from introducing artifacts that could be misinterpreted as biological plasticity.

**Real-Time Latency.** To engage BNNs in meaningful closed-loop tasks, the system must operate within the causal time window of the biological network. This requires a 'round-trip' latency (i.e., reading a spike, computing a response, and delivering stimulation) that rivals or exceeds natural synaptic delays. The CL API allows the user to request closed-loop iteration frequencies up to and including the underlying sample rate of the hardware. The maximum jitter-free iteration frequency is implementation specific and depends on the execution environment. By constraining latency and enforcing deterministic temporal semantics, a compliant implementation enables algorithms to interact with neural dynamics in real-time, effectively functioning as artificial nodes within the biological network rather than delayed observers.

## 5 The CL API Contract

In this paper, we use the term programming contract to describe a precise agreement between a software interface and its users about how operations will behave. Rather than describing how a system is implemented internally, a programming contract specifies the guarantees that users may rely on – such as timing, ordering, synchronization, and failure behavior – and the assumptions under which those guarantees hold. The key points are documented in Table 1. This approach allows experimental logic to be written against well-defined semantics, independent of the underlying hardware or software mechanisms used to realize them.

*Table 1:* Key areas and associated guarantees, assumptions, and constraints of the CL API contract.

| Item | Contract area | Guarantees, assumptions, and constraints |
|---|---|---|
| 5.1 | **Real-time closed-loop execution with defined temporal semantics** | **Real-time** refers to *defined temporal correctness*, not merely low average latency. Users may assume:<br><br>(i) operations execute with bounded latency suitable for closed-loop control;<br>(ii) temporal behavior is defined at the API level, including the meaning of timestamps, ordering, and completion;<br>(iii) transitions between requested stimulation frequencies are seamless; and<br>(iv) correctness is preserved under concurrency and load.<br><br>Timing guarantees apply under realistic experimental conditions, including overlapping stimulation, concurrent recording, and closed-loop feedback; the API prioritizes semantic correctness over best-effort execution. |
| 5.2 | **Declarative specification of intent** | The CL API is declarative: users specify what activity should occur and in what order, subject to the API contract. In particular:<br><br>(i) recording, stimulation, and closed-loop behavior are expressed as declarative requests;<br>(ii) users do not manage low-level scheduling or I/O pumping; and<br>(iii) the system enforces ordering, timing, and correctness in accordance with the contract.<br><br>This separation allows users to focus on experimental intent while the system ensures consistent and correct execution. |
| 5.3 | **Deterministic ordering, synchronization, and start-time semantics** | The CL API provides deterministic ordering and synchronization semantics across concurrent operations. Users may assume:<br><br>(i) without synchronization, execution proceeds in parallel on each channel as soon as correctness permits;<br>(ii) operations targeting the same channel execute in a deterministic order;<br>(iii) execution may be delayed by prior commitments, but will never occur earlier than requested;<br>(iv) concurrent activity across multiple channels is coordinated via explicit synchronization primitives; and<br>(v) after a synchronization operation completes, all synchronized channels commence their subsequent queued operations simultaneously. |
| 5.4 | **Atomic admission and rollback** | All stimulation and synchronisation operations submitted through the CL API are grouped into atomic transactions: a transaction is either admitted in its entirety or not admitted at all (no partial admission).<br>Operations invoked directly via the `Neurons` object (e.g., stimulation or synchronization calls) each constitute self-contained transactions, whereas operations accumulated within a `StimPlan` are grouped and submitted as a single composite transaction.<br>Admission is subject to finite, hardware-specific execution resources (e.g., per-channel stimulation queue capacity and synchronization bookkeeping capacity). If a transaction cannot be admitted due to resource constraints, the API raises `TransactionRejected` and performs a complete rollback across all affected channels. Admission is synchronous: submission returns only after the transaction has been accepted or rejected.<br>Users may assume:<br><br>(i) transactions are atomic (fully admitted or fully rejected); and<br>(ii) rejection raises `TransactionRejected`. |

| Item | Contract area | Guarantees, assumptions, and constraints |
|---|---|---|
| 5.5 | **Semantic consistency across simulation and hardware execution** | Programs written against the CL API exhibit consistent semantics across simulation and hardware execution. Specifically: <br><br> (i) the same API calls imply the same ordering, timing, and correctness guarantees; <br> (ii) simulation and hardware differ only in physical timing resolution and latency, not in meaning; <br> (iii) simulation is deterministic and authoritative with respect to API semantics; and <br> (iv) simulation enforces API-level semantic correctness but may not enforce all hardware-specific capacity constraints, so a transaction accepted in simulation may be rejected on hardware when limits are exceeded. <br><br> This supports reproducible development, testing, and validation independently of hardware availability. |
| 5.6 | **Transparency, auditability, and reproducibility** | The CL API exposes sufficient structure for reasoning about correctness and reproducibility. Users may assume: <br><br> (i) timing, ordering, and execution outcomes are observable through API-visible timestamps, events, and return values; <br> (ii) timestamps provide a consistent basis for post hoc analysis; and <br> (iii) execution behavior can be inspected and reproduced under controlled conditions. |
| 5.7 | **Scope and non-guarantees** | The CL API does **not** guarantee: <br><br> (i) absolute start-time execution without explicit synchronization; <br> (ii) zero latency or zero jitter; <br> (iii) deterministic biological responses; <br> (iv) unlimited scaling without resource contention; <br> (v) that hardware-specific resource limits are exposed or fixed across device configurations (limits are treated as implementation details); or <br> (vi) that the simulator reproduces all hardware-specific resource limits (capacity-sensitive behavior should be validated on target hardware). |

# 6 Reference Implementation: Cortical Labs CL1

This section presents the Cortical Labs CL1 system[30] as a reference implementation of the CL API contract. The examples and measurements below are selected to demonstrate that the contract's timing, ordering, synchronization, and failure semantics can be satisfied in practice under real-time and hardware resource constraints. Refer to the API Documentation for a comprehensive guide. A key feature of the CL1 that complements the CL APIs's focus on precision is reliability. In standard experimental approaches, accuracy is paramount. However, BNNs lack a stable 'ground truth' outside of reductionist patch-clamp experiments [50]. Population-level recordings via MEAs are influenced by distance from electrodes, extracellular matrix density, and cell maturity, making absolute accuracy (e.g., precise voltage at the membrane) difficult to standardize. Therefore, the CL API prioritizes *reliability* – the consistency of the interaction – over absolute accuracy. For example, whether a stimulation pulse is exactly $1.1\,\mu A$ or $1.2\,\mu A$ is secondary to ensuring that the delivered current is identical across every trial. By guaranteeing that the system's interaction with the biology is reproducible with minimal variation, this allows the features of the CL API to be fully realized for both stimulation and recording, even in the presence of biological noise.

## 6.1 Stimulation

This subsection demonstrates the contract's guarantees around precise stimulation structure and deterministic execution (Sections 5.1, 5.2, 5.3).

Stimulation is specified by a `ChannelSet` for electrode selection, a `StimDesign` defining per-phase duration, electrical current, and polarity, and optionally a `BurstDesign` to request that stimulation repeat as a burst of events at a fixed frequency:

```python
import cl

with cl.open() as neurons:
    # We'll stimulate channels 20, 42, 51, and 60
    channel_set = cl.ChannelSet(20, 42, 51, 60)

    # With a bi-phasic 200µs -2.0µA + 200µs 2.0µA stim
    stim_design = cl.StimDesign(200, -2.0, 200, 2.0)

    # As a single stimulation event,
    neurons.stim(channel_set, stim_design)

    # or as a burst of 20 stims at 37.9 Hz.
    burst_design = cl.BurstDesign(20, 37.9)
    neurons.stim(channel_set, stim_design, burst_design)
```

Calls to `Neurons.stim(...)` can take as little as 3 µs and active stimulation will begin 80 – 120 µs after that if the requested channels are idle. The minimum 80 µs lead time is the result of a tradeoff between latency and control. In this case, control over the electrical current for every pulse within a stimulation is gained at the cost of 80 µs of latency.

Additional lead time can be requested by overriding the default `lead_time_us` value of 80 µs. This is useful for exactly specifying the time between successive stimulation events on the same channel and, as discussed in Section 6.1.4, for specifying when stimulation should occur relative to stimulation on other channels.

```python
import cl

with cl.open() as neurons:
    channel_set = cl.ChannelSet((1, 2, 3, 9, 10, 11))
    stim_design = cl.StimDesign(200, -1.3, 200, 1.3)

    # Increase the stim interval by 2.5ms each time until 100ms is reached.
    for lead_time_us in range(5_000, 100_000 + 1, 2_500):
        neurons.stim(channel_set, stim_design, lead_time_us=lead_time_us)
```

### 6.1.1 Inter-Stimulus Interval Encoding

Information can be encoded in BNNs through controlled variation of stimulation timing, including inter-stimulus intervals [51]. While it is possible to manually specify temporal spacing between individual stimuli using method parameters such as `lead_time_us`, doing so requires careful coordination to avoid unintended timing artifacts.

The CL API provides a higher-level mechanism by allowing ongoing stimulation to be atomically interrupted and replaced on specified channels. By interrupting a stimulation burst operating at one frequency and enqueuing a new sequence with a different frequency, users can transition stimulation patterns without introducing spurious or uncontrolled timing gaps. The CL1 system guarantees that such interruptions take effect only at well-defined stimulus boundaries, ensuring that no unintended transitional inter-stimulus interval is introduced between successive stimulation regimes.

```
import cl
import random

with cl.open() as neurons:
    channels    = cl.ChannelSet(range(6, 64, 8))
    stim_design = cl.StimDesign(160, -1, 160, 1)

    # For 5 seconds, change the stim frequency every 250ms (4 times a second)
    for tick in neurons.loop(ticks_per_second=4, stop_after_ticks=20):
        new_frequency = random.uniform(4, 200) # range [4, 200] Hz
        neurons.interrupt_then_stim(
            channels,
            stim_design,
            cl.BurstDesign(1000, new_frequency))

    # Interrupt the final burst of 1000.
    neurons.interrupt(channels)
```

### 6.1.2 Burst Frequency Quantization

As the CL1 operates with a fixed stimulation base frequency of 50 kHz, burst frequencies are approximated using the nearest frequency representable by an integer multiple of the minimum stimulation period (20 µs). Within the CL1 simulation frequency limit of 200 Hz, the maximum error deviation from a requested frequency is 0.2 % (see Figure 1).

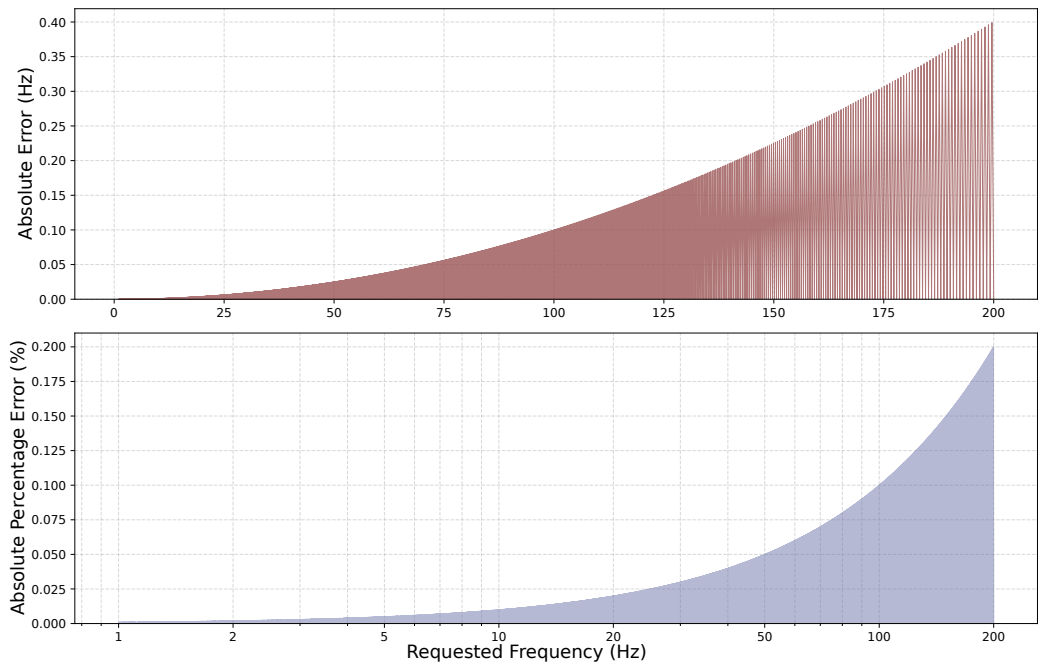

*Figure 1:* Absolute error (top) and percentage error (bottom) to the closest possible frequency, from a requested frequency (x-axis).

### 6.1.3 Channel Synchronization

Each channel (electrode) maintains an independent stimulation queue. When a queue is empty, a newly requested operation begins as soon as it becomes visible to the CL1 timing engine; when a queue is not empty, newly requested operations are appended to the end of the queue. In the absence of additional coordination mechanisms, this per-channel queuing model makes precise temporal alignment of stimulation across multiple channels non-trivial.

To address this, the `Neurons.sync(...)` operation provides an explicit coordination barrier across channels. A synchronization request specifies a set of channels and delays the execution of subsequently queued stimulation on those channels until all prior activity on each has completed. This ensures that stimulation following a synchronization point begins simultaneously across the specified channels, regardless of differences in per-channel queue depth or execution history.

Stimulation methods such as `Neurons.stim(...)` automatically introduce synchronization when operating on a channel set containing more than one channel. Explicit use of `Neurons.sync(...)` is therefore required only when coordination is needed between otherwise unrelated channels.

As any requirement for coordinated simultaneous stimulation is more reliably addressed with a `StimPlan` (see 6.1.4), the primary use of `Neurons.sync(...)` is to delay a new stimulation until previous activity on synchronized channels has completed.

Figure 2 (Scenario A) illustrates two independent, non-overlapping `ChannelSet` instances, each associated with its own `StimDesign`. Because the stimulation queues for all channels are initially empty, stimulation on all channels begins simultaneously. Figure 2 (Scenario B) introduces a call to `Neurons.sync(...)` over the combined channel sets, which causes the second stimulation to wait for the first to finish before starting.

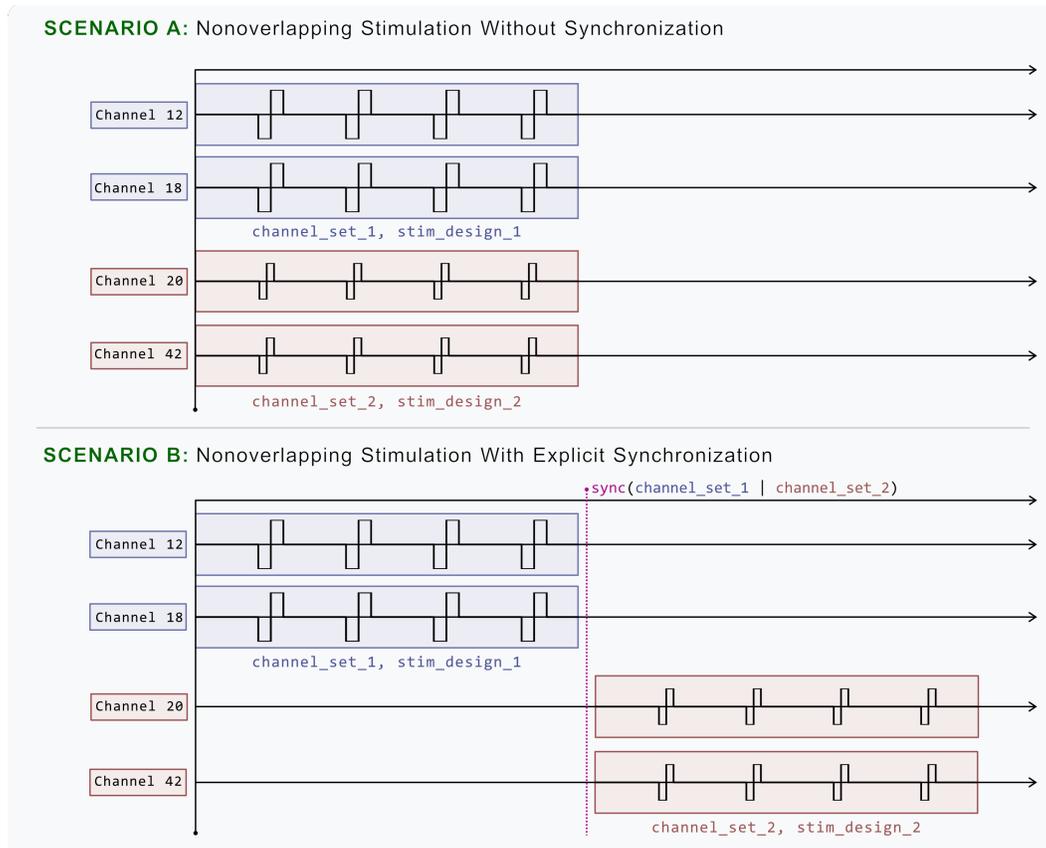

*Figure 2:* Queued stimulation scenarios with and without `Neurons.sync(...)`. Channels highlighted in blue are part of `channel_set_1`, whereas channels highlighted in red are part of `channel_set_2`. In Scenario A (without synchronization), all stimulation operations start simultaneously. In Scenario B, `Neurons.sync(channel_set_1 | channel_set_2)` is invoked between `Neurons.stim(...)` calls, resulting in stimulation of `channel_set2` waiting until the completion of previous stimulation in `channel_set_1`.

### 6.1.4 Stimulation Plans

Even the simplest call to `Neurons.stim(...)` takes approximately 3 µs to execute in user space. On the CL1, newly submitted operations become visible to the hardware timing engine at fixed 40 µs intervals, corresponding to frames of raw ADC samples. As a result, stimulation triggered by successive calls to `Neurons.stim(...)` may not begin at exactly the same time, since those calls may be admitted in different timing frames.

Stimulation plans allow multiple related stimulation requests to be admitted transactionally (all, or none in the event of failure, never partial) and to become visible simultaneously. When combined with inter-channel synchronization (6.1.3), stimulation plans provide precise control over which stimulation is active on each channel and when that activity begins relative to other activity within the same plan.

The `StimPlan` class supports the same `stim(...)` and `sync(...)` methods as the `Neurons` class, with the key distinction that operations are accumulated but not made visible until `StimPlan.run(...)` is invoked. A stimulation plan may be executed multiple times, once per invocation of `StimPlan.run(...)`. Execution may also be scheduled for a specific future timestamp via `StimPlan.run(at_timestamp=...)`.

While stimulation plans support interruption of activity on specified channels, this capability is expressed declaratively via the `StimPlan.channels_to_interrupt` property rather than through imperative interruption calls such as `interrupt(...)` or `interrupt_then_stim(...)`. This reflects the fact that a stimulation plan can interrupt existing activity only at the moment it becomes visible, ensuring that interruption and subsequent stimulation occur atomically at a well-defined execution boundary.

```python
import cl

# First set of channels
channel_set_1 = cl.ChannelSet(20, 42, 51, 60)
stim_design_1  = cl.StimDesign(200, -2.0, 200, 2.0)
burst_design_1 = cl.BurstDesign(20, 40)

# Second set of channels
channel_set_2 = cl.ChannelSet(2, 6, 12, 18)
stim_design_2  = cl.StimDesign(100, -1.5, 100, 1.5)
burst_design_2 = cl.BurstDesign(10, 20)

# Create a combined set of all channels used above
all_target_channels = channel_set_1 | channel_set_2

with cl.open() as neurons:

    # Create a transactional stimulation plan
    stim_plan = neurons.create_stim_plan()

    # Interrupt our target channels in case any are in use
    stim_plan.channels_to_interrupt = all_target_channels

    # Block target channels until the interruption process is
    # complete and all channels are ready.
    stim_plan.sync(all_target_channels)

    # We want to stimulate each channel group at the same time.
    # This is guaranteed as the channel sets do not overlap.
    stim_plan.stim(channel_set_1, stim_design_1, burst_design_1)
    stim_plan.stim(channel_set_2, stim_design_2, burst_design_2)

    # Submit our plan for asap execution.
    stim_plan.run()

    # Run the same plan again two seconds later.
    stim_plan.run(neurons.timestamp() + (2 * neurons.get_frames_per_second()))
```

In the above example, the two `ChannelSet` instances do not overlap, and the corresponding stimulation sequences therefore commence simultaneously across their respective channels. Each `StimDesign` applies exclusively to its associated `ChannelSet`. Consequently, channels 2, 6, 12 and 18 are stimulated with biphasic pulses of $1.5\,\mu A$ and a pulse width of $100\,\mu s$, while channels 20, 42, 51 and 60 are stimulated with biphasic pulses of $2\,\mu A$ and a pulse width of $200\,\mu s$. Because each group is associated with a distinct `BurstDesign`, the two channel sets operate at different stimulation frequencies.

### 6.1.5 Detecting and Reacting to Spikes in Real-Time

The `Neurons.loop(...)` method provides a time-driven execution context in which detected action potentials (spikes) are surfaced to user code at a fixed iteration frequency of up to 25 kHz. Each iteration of the loop corresponds to a well-defined timing frame in the underlying system, allowing user logic to observe neural activity and issue responses with bounded and repeatable latency.

Within each loop iteration, the system exposes all spikes detected since the previous iteration via the `tick.analysis.spikes` list. This enables closed-loop algorithms to react to neural activity at timescales comparable to the underlying sampling and stimulation hardware, without requiring users to manage low-level buffering, polling, or synchronization.

```python
import cl

with cl.open() as neurons:
    # Loop 1000 times per second for 5 seconds
    for tick in neurons.loop(ticks_per_second=1000, stop_after_seconds=5):
        # Loop through each detected spike object
        for spike in tick.analysis.spikes:
            # Print out the spike object
            print(spike)
```

```
Spike(timestamp=2707168587, channel=52)
Spike(timestamp=2707168645, channel=58)
Spike(timestamp=2707168737, channel=32)
Spike(timestamp=2707168855, channel=45)
Spike(timestamp=2707168987, channel=43)
...
Spike(timestamp=2707291908, channel=58)
Spike(timestamp=2707292040, channel=21)
Spike(timestamp=2707292056, channel=60)
Spike(timestamp=2707292259, channel=52)
Spike(timestamp=2707292898, channel=62)
```

A `Spike` object is created for each detected action potential and includes channel and timestamp metadata, along with the ability to retrieve a short window of raw samples surrounding the detection event. The list of spikes provided in each iteration reflects all detections that occurred since the previous loop iteration, ensuring that no events are missed even when multiple spikes occur within a single timing frame.

`Spike` objects expose the following properties:

| Property | Data |
| --- | --- |
| `timestamp` | Timestamp of the sample that triggered the detection of the spike |
| `channel` | Which channel the spike was detected on |
| `samples` | NumPy array of 75 floating point µV sample values around `timestamp` |

The `samples` array is mean-centered per channel at the time of detection, removing the instantaneous channel baseline while preserving relative waveform shape. Sample values are converted to physical units and provided in floating point microvolts (µV). The array contains 1 ms of samples preceding the detection timestamp and 2 ms of samples beginning at the detection timestamp and extending forward.

Because spike processing logic commonly executes well within 2 ms of detection, sample data is populated lazily and loaded only when requested. Accessing `Spike.samples` before the full post-detection window has been acquired will block until the required data becomes available, potentially blocking execution by up to 2 ms. To bound resource usage, lazy loading is only supported for a limited time window; to succeed, `Spike.samples` must have been first accessed within an approximate 5 second window after spike detection.

The `Neurons.loop(...)` also exposes `tick.analysis.stims`, which contains a list of `Stim` objects that serve as a record of stimulation that began during the previous tick. The following 25 kHz loop stimulates in response to each detected spike, then collates and later prints detected spikes and stims:

```python
import cl

with cl.open() as neurons:
    # Loop at 25 kHz for 25000 ticks, respond to spikes
    # with a stim on the same channel, and collect all
    # detected spikes and stims.
    spikes = []
    stims  = []
    stim_design = cl.StimDesign(160, -1, 160, 1)
    for tick in neurons.loop(ticks_per_second=25000, stop_after_ticks=25000):
        # Respond to each detected spike with a stim on the same channel.
        for spike in tick.analysis.spikes:
            neurons.stim(spike.channel, stim_design)
        # Collect the spikes and stims from the previous tick
        spikes.extend(tick.analysis.spikes)
        stims.extend(tick.analysis.stims)

for spike_or_stim in sorted(spikes + stims, key=lambda x: x.timestamp):
    print(spike_or_stim)
```

```
Spike(timestamp=1195060728, channel=58)
Stim(timestamp=1195060733, channel=58)
Spike(timestamp=1195061224, channel=5)
Stim(timestamp=1195061229, channel=5)
Spike(timestamp=1195061710, channel=54)
Stim(timestamp=1195061715, channel=54)
...
Spike(timestamp=1195084744, channel=42)
Stim(timestamp=1195084749, channel=42)
Spike(timestamp=1195085149, channel=54)
Stim(timestamp=1195085154, channel=54)
Spike(timestamp=1195085546, channel=36)
Stim(timestamp=1195085550, channel=36)
```

## 6.2 Jitter Detection

The CL API Contract (Section 5) defines explicit guarantees regarding temporal correctness and bounded execution latency. For closed-loop neurocomputing experiments, it is essential that analysis and response logic execute within predictable and well-defined time windows; otherwise, apparent experimental effects may be confounded by uncontrolled timing variability.

The preceding example demonstrated that neural spikes can be detected and exposed to Python-level logic at iteration frequencies of up to 25 kHz. At this rate, each loop iteration corresponds to a 40 µs timing frame. After accounting for fixed system overheads, the user-level loop body has an effective execution budget of approximately 33 µs to 34 µs. This bound can be demonstrated empirically.

```python
import cl

from time import time_ns

with cl.open() as neurons:
    for tick in neurons.loop(ticks_per_second=25000, stop_after_ticks=25000):
        # Consume approximately 33 µs of execution time
        wait_until_ns = time_ns() + 33_000
        while time_ns() < wait_until_ns:
            pass

print('Done')
```

```
Done.
```

When the execution time of a loop iteration exceeds the available budget, the system detects that the loop has fallen behind the hardware timing engine and raises a `TimeoutError`. This behavior enforces the temporal guarantees of the API contract by preventing silent accumulation of jitter.

```
import cl

from time import time_ns

with cl.open() as neurons:
    for tick in neurons.loop(ticks_per_second=25000, stop_after_ticks=25000):
        # Consume approximately 35 µs, exceeding the 25 kHz per-iteration budget
        wait_until_ns = time_ns() + 35_000
        while time_ns() < wait_until_ns:
            pass

print('Done')
```

```
---------------------------------------------------------------------------
TimeoutError                              Traceback (most recent call last)
Cell In[64], line 6
      3 from time import time_ns
      5 with cl.open() as neurons:
----> 6     for tick in neurons.loop(ticks_per_second=25000, stop_after_ticks=25000):
      7         # Consume approximately 35 µs, exceeding the 25 kHz per-iteration budget
      8         wait_until_ns = time_ns() + 35_000
      9         while time_ns() < wait_until_ns:

TimeoutError: Loop fell behind by 1 frame (40 µs) when entering the 8th
iteration. Jitter tolerance is currently set to 0 frames. Ideally - optimise
the worst-case performance of your loop body. You may also adjust the jitter
tolerance via jitter_tolerance_frames=1, or ignore jitter entirely via
ignore_jitter=True.
```

It is important to note that iterations of `Neurons.loop(...)` are scheduled at fixed, regular intervals. The jitter detection subsystem does not evaluate average execution time; instead, it enforces a strict per-iteration deadline relative to the availability of newly sampled electrode data. Each loop body must complete within its allotted timing frame, or the iteration is considered to have missed its deadline.

## 6.3 Jitter Recovery and Temporal Synchronization

In some closed-loop experiments, it is desirable to temporarily relax enforcement of per-iteration execution deadlines. For example, an application may need to perform a blocking operation – such as communicating with a remote service – between timing-critical sections of a control loop. While such behavior can always be implemented by exiting a `Neurons.loop(...)` context and later starting a new one, the CL API also provides a structured mechanism for handling controlled timing disruptions within a single loop.

The `Loop.recover_from_jitter(...)` method allows a loop to tolerate a bounded lapse in execution without triggering a jitter error, while preserving well-defined temporal semantics. This mechanism is explicitly opt-in and applies only when invoked by the user.

When `recover_from_jitter(...)` is called, the `Loop` instance enters a recovery phase in which the standard loop body is temporarily bypassed. During this phase, the internal iteration counter advances rapidly to realign the logical tick index with the underlying frame clock. Data associated with skipped iterations may optionally be forwarded to a user-defined callback (`handle_recovery_tick`), allowing applications to maintain state consistency while temporal alignment is restored.

```
def recover_from_jitter(
    self,
    handle_recovery_tick: Callable[[LoopTick], None] | None = None,
    timeout_seconds: float = 5.0
) -> None
```

Successful temporal resynchronization is subject to several operational constraints. The optional `handle_recovery_tick` callback must remain computationally lightweight; if its execution time approaches the loop's nominal tick interval, the recovery process will be unable to close the latency gap and resynchronization will fail.

The `timeout_seconds` parameter acts as a watchdog for the recovery process, preventing the system from entering an unbounded recovery state if the accumulated processing lag consistently exceeds the rate at which logical iterations can be advanced.

Finally, jitter recovery is limited by a finite retrospective data window. Recovery can succeed only while the system retains sufficient historical data to reconcile skipped iterations, which in the current CL1 implementation corresponds to a window of approximately 5 s.

The following example illustrates controlled handling of a transient blocking operation within a closed-loop execution. During the third iteration, the loop body intentionally exceeds its execution budget, causing subsequent iterations to fall behind the real-time schedule. The jitter recovery mechanism is then invoked, at which point the loop enters a recovery phase in which delayed iterations are skipped, associated callbacks are invoked, and the logical iteration index is advanced to reestablish temporal alignment. Normal execution resumes once the loop has caught up, in this case at iteration 6.

```python
import cl

from time import sleep

TICKS_PER_SECOND = 100
SECONDS_PER_TICK = 1 / TICKS_PER_SECOND

def recovery_callback(tick):
    # Lightweight processing for skipped ticks
    print(f"Iteration: {tick.iteration} skipped during recovery")

with cl.open() as neurons:
    for tick in neurons.loop(TICKS_PER_SECOND, stop_after_ticks=10):
        print(f"Iteration: {tick.iteration} handled normally")

        # On the third iteration,
        if tick.iteration == 2:
            # exceed our time budget,
            sleep(3 * SECONDS_PER_TICK)
            # but let the system know this was expected.
            tick.loop.recover_from_jitter(handle_recovery_tick=recovery_callback)
```

```
Iteration: 0 handled normally
Iteration: 1 handled normally
Iteration: 2 handled normally
Iteration: 3 skipped during recovery
Iteration: 4 skipped during recovery
Iteration: 5 skipped during recovery
Iteration: 6 handled normally
Iteration: 7 handled normally
Iteration: 8 handled normally
Iteration: 9 handled normally
```

## 6.4 Recording

This subsection demonstrates that the observability and reproducibility guarantees of the CL API contract are satisfied without requiring application logic to manage I/O concerns.

The CL1 recording subsystem provides a structured mechanism for producing HDF5 recordings that capture detected spikes, stimulation events, application-defined time series data, and a complete record of raw electrode samples. An application initiates recording via a single API call; by default, all spikes, stimulation events, and sampled data frames are captured continuously until recording is stopped. Crucially, the application is not responsible for explicitly pumping or scheduling the recording I/O path, which is handled transparently by the system.

```python
import cl

from .my_game import game

with cl.open() as neurons:
    recording = neurons.record()

    # Recording has started; application logic may perform
    # arbitrary blocking foreground work.
    game.run(neurons)

    recording.stop()

print(recording.file['path'])
```

```
/data/recordings/2026-02-07_00-04-05.213+00-00_recording.h5
```

The recording system further supports precise control over recording duration, selective inclusion of data modalities, and bounded retrospective capture from the recent past, enabling reproducible experiment reconstruction without entangling recording concerns with application control flow.

### 6.4.1 Data Streams

Data streams provide a mechanism for applications to publish named, timestamped streams of structured data that are incorporated into recordings and made available for live visualization. Each data stream is identified by name and associates each entry with an explicit timestamp, allowing application-level state to be aligned precisely with neural activity and stimulation events. Data streams are created via `Neurons.create_data_stream(...)`.

```python
import cl
import numpy

with cl.open() as neurons:
    # Create a named data stream - by default, it will be added to any active or future recordings.
    data_stream = \
        neurons.create_data_stream(
            name       = 'example_data_stream',
            attributes = { 'score': 0, 'another_attribute': [0, 1, 2, 3] }
        )

    # Start a recording
    recording = neurons.record(stop_after_seconds=1)
```

Data stream entries are well suited to representing time-varying application state, such as the $(x, y)$ position of an object in a closed-loop task or intermediate values produced by an analysis pipeline. Each entry must be associated with a unique, strictly increasing timestamp. Entry payloads may be arbitrary Python objects, including `dict`, `tuple`, `list`, scalar numeric types, or NumPy `NDArray` instances.

Attributes may be defined to describe or summarize the state of a data stream. Unlike entries, attributes store only their most recent value.

After data stream creation, entries may be appended as follows:

```python
import cl
import numpy

with cl.open() as neurons:
     # Create a named data stream - by default, it will be added
     # to any active or future recordings.
    data_stream = \
        neurons.create_data_stream(
            name='example_data_stream',
            attributes={ 'score': 0, 'another_attribute': [0, 1, 2, 3] })

    # Start a recording
    recording = neurons.record(stop_after_seconds=1)
    timestamp = neurons.timestamp()  # get current timestamp

    # Add some data stream entries with unique, ascending timestamps:
    data_stream.append(timestamp + 0, { 'arbitrary': 'data' })
    data_stream.append(timestamp + 1, ['of', 'arbitrary', 'size'])
    data_stream.append(timestamp + 2, 'and type.')
    data_stream.append(
        timestamp + 3,
        numpy.array([2**64 - 1, 2**64 - 2, 2**64 - 3], dtype=numpy.uint64)
    )

    # Update attributes
    data_stream.set_attribute('score', 1)
    data_stream.update_attributes({ 'score': 2, 'new_attribute': 9.9 })

    recording.wait_until_stopped()

print(recording)
```

```
Recording(status=stopped, start_timestamp=1449386250, end_timestamp=1449411250, duration=25000,
↪   file_path=/data/recordings/2026-02-11_01-25-37.188+00-00_recording.h5)
```

Data stream entries may be appended at any time, independent of whether a recording is currently active. When a recording is active, entries are captured with their associated timestamps; otherwise, they remain available for subsequent recordings or live visualization.

## 6.5 Loading Recordings

Recordings produced by the CL1 system are stored in a structured HDF5 format that enables post hoc inspection, analysis, and reproducibility. The `RecordingView` interface provides programmatic access to recorded data without requiring knowledge of the underlying file layout.

```python
from cl import RecordingView

# Load a recording file from CL1 using a RecordingView
# Files are timestampted in a `YYYY_MM_DD_HH_MM_SS` format
recording = RecordingView("/data/recordings/2026-02-11_01-25-37.188+00-00_recording.h5")

# Access core datasets
print(recording.samples)       # Raw voltage samples (T × C array)
print(recording.spikes)        # Detected spike events with metadata
print(recording.stims)         # Stimulus timestamps and channels
print(recording.attributes)    # Global recording metadata
print(recording.data_streams)  # Additional experiment streams (e.g., gamestate)

# Example: Count total spikes
num_spikes = len(recording.spikes)
print(f"Detected spikes: {num_spikes}")

# Close file when done
recording.close()
```

```
/samples (EArray(np.int64(25000), np.int64(64))) ''
/spikes (Table(np.int64(97),)) np.str_('')
/stims (Table(np.int64(0),)) np.str_('')
AttributesView:
{
    'app_info': {'id': 'cl-jupyter', 'name': 'CL API Whitepaper', 'path': '/data/notebooks/CL API
    ↪   Whitepaper.ipynb'}
    'application': {}
    'cell_batch_id': ''
    'channel_count': 64
    'chip_id': ''
    'created_localtime': '2026-02-11T01:25:37.193617+00:00'
    'created_utc': '2026-02-11T01:25:37.193617+00:00'
    'duration_frames': 25000
    'duration_seconds': 1.0
    'end_timestamp': 1449411250
    'ended_localtime': '2026-02-11T01:25:38.209845+00:00'
    'ended_utc': '2026-02-11T01:25:38.209845+00:00'
    'file_format': {'version': 2, 'stim_and_spike_timestamps_relative_to_start': True}
    'frames_per_second': 25000
    'git_branch': 'dqh/doc-stim'
    'git_hash': 'bc044040bb53f5e0b395ad982f01c7739c42c807'
    'git_status': 'clean'
    'git_tag': ''
    'git_tags': ['']
    'hostname': 'CL1-2507-8'
    'plugin': {}
    'project_id': ''
    'sampling_frequency': 25000
    'start_timestamp': 1449386250
    'system_id': 'CL1-2507-8'
    'uV_per_sample_unit': 0.19499999284744263
}
Data Streams:
    example_data_stream
Detected spikes: 97
```

The metadata stored alongside each recording captures sufficient contextual information to support experiment reconstruction and auditability, including timing parameters, hardware identifiers, software versioning, and application-level annotations.

All timestamps stored within a recording—including spike times, stimulation events, and data stream entries—are expressed as monotonically increasing frame indices relative to the start of the recording. This frame-relative time base provides a stable and self-contained reference for post hoc analysis, allowing recorded events to be aligned and compared without reliance on absolute or wall-clock time.

### 6.5.1 Loading Data Streams

Application-defined data streams stored within a recording can be accessed in a similar manner. The following example demonstrates how to iterate over the entries of a previously recorded data stream (`example_data_stream`), preserving the original timestamps associated with each entry.

```python
from cl import RecordingView

recording = RecordingView("/data/recordings/2026-02-11_01-25-37.188+00-00_recording.h5")

for timestamp, data in recording.data_streams.example_data_stream.items():
    print(timestamp, data)
```

```
4461 {'arbitrary': 'data'}
4462 ['of', 'arbitrary', 'size']
4463 and type.
4464 [18446744073709551615 18446744073709551614 18446744073709551613]
```

This interface allows application state, derived quantities, and experimental annotations to be analyzed alongside neural recordings using a common time base, without requiring custom parsing or synchronization logic.

# 7 Conclusion

As work exploring *in vitro* neural cultures continues to advance, the need for reliable tools that are widely accessible becomes critical. This paper has presented the CL API as a fundamental abstraction for real-time interaction with biological neural networks (BNNs). The purpose of this API is not merely to expose underlying hardware mechanisms to a high-level language, but also to define a clear and reliable *contract* governing the execution, timing, and coordination of neural recording and stimulation. The resulting CL API can allow rapid iteration and development of tightly controlled electrophysiological real-time closed-loop environments for BNN.

This contract makes explicit the guarantees that users of the CL API may rely upon, regardless of the underlying platform. These commitments define the semantic boundary of the system. By formalizing these guarantees, the CL API decouples the experimental logic from the implementation details. For the neurocomputing researcher, this means that the 'black box' of hardware execution is replaced by a transparent, predictable interface. The adoption of a rigorous execution contract is a necessary step towards addressing the reproducibility crisis in the field, transforming BNN research from bespoke, hardware-dependent demonstrations into a rigorous, reproducible, and scalable discipline.

Equally important, these performance benefits are achieved in coordination with ease-of-use, with significant modularity and transparency. Democratizing access to BNNs via the CL API empowers a broader spectrum of scientists and engineers. This inclusivity not only enhances reproducibility but also catalyzes novel developments, accelerates discovery, and fosters multidisciplinary collaboration. Ultimately, the CL API will be a powerful tool that aims to accelerate the rate of progress when undertaking electrophysiological interactions with BNN.

# References

[1] Brett J. Kagan et al. "Neurons Embodied in a Virtual World: Evidence for Organoid Ethics?" In: *AJOB Neuroscience* 13.2 (Apr. 2022), pp. 114–117. ISSN: 2150-7740. DOI: 10.1080/21507740.2022.2048731.

[2] Brett J. Kagan et al. "Scientific communication and the semantics of sentience". en. In: *Neuron* 111.5 (Mar. 2023), pp. 606–607. ISSN: 08966273. DOI: 10.1016/j.neuron.2023.02.008.

[3] Brett J. Kagan et al. "Harnessing Intelligence from Brain Cells In Vitro". In: *The Neuroscientist* 31.5 (Oct. 2025), pp. 536–555. ISSN: 1073-8584. DOI: 10.1177/10738584251321438.

[4] Lena Smirnova et al. "Organoid intelligence (OI): the new frontier in biocomputing and intelligence-in-a-dish". In: *Frontiers in Science* Volume 1 - 2023 (2023). ISSN: 2813-6330. DOI: 10.3389/fsci.2023.1017235.

[5] Itzy E. Morales Pantoja et al. "First Organoid Intelligence (OI) workshop to form an OI community". en. In: *Frontiers in Artificial Intelligence* 6 (Feb. 2023), p. 1116870. ISSN: 2624-8212. DOI: 10.3389/frai.2023.1116870.

[6] Xiaotian Zhang et al. "Mind In Vitro Platforms: Versatile, Scalable, Robust, and Open Solutions to Interfacing with Living Neurons". en. In: *Advanced Science* 11.11 (Mar. 2024), p. 2306826. ISSN: 2198-3844, 2198-3844. DOI: 10.1002/advs.202306826.

[7] Thomas Hartung et al. "The Baltimore declaration toward the exploration of organoid intelligence". en. In: *Frontiers in Science* 1 (Feb. 2023), p. 1068159. ISSN: 2813-6330. DOI: 10.3389/fsci.2023.1068159.

[8] Fred D. Jordan et al. "Open and remotely accessible Neuroplatform for research in wetware computing". en. In: *Frontiers in Artificial Intelligence* 7 (May 2024), p. 1376042. ISSN: 2624-8212. DOI: 10.3389/frai.2024.1376042.

[9] Emre O. Neftci and Bruno B. Averbeck. "Reinforcement learning in artificial and biological systems". en. In: *Nature Machine Intelligence* 1.3 (Mar. 2019), pp. 133–143. ISSN: 2522-5839. DOI: 10.1038/s42256-019-0025-4.

[10] Moein Khajehnejad et al. "Dynamic Network Plasticity and Sample Efficiency in Neural Cultures: A Comparison with Deep Learning". en. In: *Cyborg and Bionic Systems* (June 2025). ISSN: 2692-7632. DOI: 10.34133/cbsystems.0336.

[11] Brett J. Kagan et al. "In vitro neurons learn and exhibit sentience when embodied in a simulated game-world". In: *Neuron* 110.23 (Dec. 2022), 3952–3969.e8. ISSN: 0896-6273. DOI: 10.1016/j.neuron.2022.09.001.

[12] Mattia D'Andola et al. "Bistability, Causality, and Complexity in Cortical Networks: An In Vitro Perturbational Study". en. In: *Cerebral Cortex* 28.7 (July 2018), pp. 2233–2242. ISSN: 1047-3211, 1460-2199. DOI: 10.1093/cercor/bhx122.

[13] David Beniaguev, Idan Segev, and Michael London. "Single cortical neurons as deep artificial neural networks". en. In: *Neuron* 109.17 (Sept. 2021), 2727–2739.e3. ISSN: 08966273. DOI: 10.1016/j.neuron.2021.07.002.

[14] Suhas Kumar, R. Stanley Williams, and Ziwen Wang. "Third-order nanocircuit elements for neuromorphic engineering". en. In: *Nature* 585.7826 (Sept. 2020), pp. 518–523. ISSN: 0028-0836, 1476-4687. DOI: 10.1038/s41586-020-2735-5.

[15] Brett J. Kagan et al. "The technology, opportunities, and challenges of Synthetic Biological Intelligence". In: *Biotechnology Advances* 68 (Nov. 2023), p. 108233. ISSN: 0734-9750. DOI: 10.1016/j.biotechadv.2023.108233.

[16] Christian Hugo Hoffmann. "A philosophical view on singularity and strong AI". en. In: *AI & SOCIETY* 38.4 (Aug. 2023), pp. 1697–1714. ISSN: 0951-5666, 1435-5655. DOI: 10.1007/s00146-021-01327-5.

[17] Hao Wang et al. "NeuroD4 converts glioblastoma cells into neuron-like cells through the SLC7A11-GSH-GPX4 antioxidant axis". en. In: *Cell Death Discovery* 9.1 (Aug. 2023), p. 297. ISSN: 2058-7716. DOI: 10.1038/s41420-023-01595-8.

[18] Forough Habibollahi et al. "Critical dynamics arise during structured information presentation within embodied in vitro neuronal networks". en. In: *Nature Communications* 14.1 (Aug. 2023), p. 5287. ISSN: 2041-1723. DOI: 10.1038/s41467-023-41020-3.

[19] Moein Khajehnejad et al. "On Complex Network Dynamics of an In-Vitro Neuronal System during Rest and Gameplay". English. In: *NeurIPS 2023 Generative AI and Biology (GenBio) Workshop*. Vol. 2023. 1. NeurIPS, 2023, p. 1.

[20] Douglas J Bakkum, Zenas C Chao, and Steve M Potter. "Spatio-temporal electrical stimuli shape behavior of an embodied cortical network in a goal-directed learning task". In: *Journal of neural engineering* 5.3 (2008), p. 310. ISSN: 1741-2552.

[21] Jacopo Tessadori et al. "Modular Neuronal Assemblies Embodied in a Closed-Loop Environment: Toward Future Integration of Brains and Machines". en. In: *Frontiers in Neural Circuits* 6 (2012). ISSN: 1662-5110. DOI: 10.3389/fncir.2012.00099.

[22] Y. Jimbo, H.P.C. Robinson, and A. Kawana. "Strengthening of synchronized activity by tetanic stimulation in cortical cultures: application of planar electrode arrays". en. In: *IEEE Transactions on Biomedical Engineering* 45.11 (Nov. 1998), pp. 1297–1304. ISSN: 00189294. DOI: 10.1109/10.725326.

[23] Goded Shahaf and Shimon Marom. "Learning in Networks of Cortical Neurons". en. In: *The Journal of Neuroscience* 21.22 (Nov. 2001), pp. 8782–8788. ISSN: 0270-6474, 1529-2401. DOI: 10.1523/JNEUROSCI.21-22-08782.2001.

[24] F. Heer et al. "CMOS microelectrode array for bidirectional interaction with neuronal networks". en. In: *Proceedings of the 31st European Solid-State Circuits Conference, 2005. ESSCIRC 2005*. Grenoble, France: IEEE, 2005, pp. 335–338. ISBN: 978-0-7803-9205-2. DOI: 10.1109/ESSCIR.2005.1541628.

[25] M Jenkner et al. "Cell-based CMOS sensor and actuator arrays". en. In: 39.12 (2004), p. 7.

[26] Y. Jimbo et al. "A system for MEA-based multisite stimulation". In: *IEEE Transactions on Biomedical Engineering* 50.2 (Feb. 2003), pp. 241–248. ISSN: 1558-2531. DOI: 10.1109/TBME.2002.805470.

[27] Anindita Sarkar et al. "Efficient Generation of CA3 Neurons from Human Pluripotent Stem Cells Enables Modeling of Hippocampal Connectivity In Vitro". English. In: *Cell Stem Cell* 22.5 (May 2018), 684–697.e9. ISSN: 1934-5909, 1875-9777. DOI: 10.1016/j.stem.2018.04.009.

[28] Yichen Shi, Peter Kirwan, and Frederick J Livesey. "Directed differentiation of human pluripotent stem cells to cerebral cortex neurons and neural networks". en. In: *Nature Protocols* 7.10 (Oct. 2012), pp. 1836–1846. ISSN: 1754-2189, 1750-2799. DOI: 10.1038/nprot.2012.116.

[29] Madeline A. Lancaster et al. "Cerebral organoids model human brain development and microcephaly". en. In: *Nature* 501.7467 (Sept. 2013), pp. 373–379. ISSN: 0028-0836, 1476-4687. DOI: 10.1038/nature12517.

[30] Brett J. Kagan. "The CL1 as a platform technology to leverage biological neural system functions". en. In: *Nature Reviews Bioengineering* (July 2025). ISSN: 2731-6092. DOI: 10.1038/s44222-025-00340-3.

[31] Edward S. Boyden et al. "Millisecond-timescale, genetically targeted optical control of neural activity". en. In: *Nature Neuroscience* 8.9 (Sept. 2005), pp. 1263–1268. ISSN: 1546-1726. DOI: 10.1038/nn1525.

[32] Valentina Emiliani et al. "Optogenetics for light control of biological systems". en. In: *Nature Reviews Methods Primers* 2.1 (July 2022), p. 55. ISSN: 2662-8449. DOI: 10.1038/s43586-022-00136-4.

[33] Yulia Mourzina et al. "Spatially resolved non-invasive chemical stimulation for modulation of signalling in reconstructed neuronal networks". In: *Journal of The Royal Society Interface* 3.7 (Nov. 2005), pp. 333–343. ISSN: 1742-5689. DOI: 10.1098/rsif.2005.0099.

[34] Andrea Brüggemann et al. "Planar patch clamp: advances in electrophysiology". In: *Potassium Channels: Methods and Protocols* (2009), pp. 165–176. ISSN: 1934115657.

[35] R. Becket Ebitz and Benjamin Y. Hayden. "The population doctrine in cognitive neuroscience". en. In: *Neuron* (Aug. 2021), S0896627321005213. ISSN: 08966273. DOI: 10.1016/j.neuron.2021.07.011.

[36] Takuya Isomura, Kiyoshi Kotani, and Yasuhiko Jimbo. "Cultured Cortical Neurons Can Perform Blind Source Separation According to the Free-Energy Principle". en. In: *PLOS Computational Biology* 11.12 (Dec. 2015), e1004643. ISSN: 1553-7358. DOI: 10.1371/journal.pcbi.1004643.

[37] Hideaki Yamamoto et al. "Modular architecture facilitates noise-driven control of synchrony in neuronal networks". en. In: *Science Advances* 9.34 (Aug. 2023), eade1755. ISSN: 2375-2548. DOI: 10.1126/sciadv.ade1755.

[38] Hongwei Cai et al. "Brain organoid reservoir computing for artificial intelligence". In: *Nature Electronics* 6.12 (Dec. 2023), pp. 1032–1039. ISSN: 2520-1131. DOI: 10.1038/s41928-023-01069-w.

[39] Hyunjun Jung, Jintae Kim, and Yoonkey Nam. "Recovery of early neural spikes from stimulation electrodes using a DC-coupled low gain high resolution data acquisition system". en. In: *Journal of Neuroscience Methods* 304 (July 2018), pp. 118–125. ISSN: 0165-0270. DOI: 10.1016/j.jneumeth.2018.04.014.

[40] Danny Hsu Ko et al. "3D microelectrode arrays, pushing the bounds of sensitivity toward a generic platform for point-of-care diagnostics". en. In: *Biosensors and Bioelectronics* 227 (May 2023), p. 115154. ISSN: 09565663. DOI: 10.1016/j.bios.2023.115154.

[41] Francisco Fambrini, José Hiroki Saito, and Luis Mariano Del Val Cura. "MEA recording system circuit implementation". In: *IECON 2017 - 43rd Annual Conference of the IEEE Industrial Electronics Society*. Oct. 2017, pp. 8515–8520. DOI: 10.1109/IECON.2017.8217495.

[42] Jan Müller, Douglas J. Bakkum, and Andreas Hierlemann. "Sub-millisecond closed-loop feedback stimulation between arbitrary sets of individual neurons". en. In: *Frontiers in Neural Circuits* 6 (2013). ISSN: 1662-5110. DOI: 10.3389/fncir.2012.00121.

[43] Jonathan P. Newman et al. "Closed-Loop, Multichannel Experimentation Using the Open-Source NeuroRighter Electrophysiology Platform". en. In: *Frontiers in Neural Circuits* 6 (2013). ISSN: 1662-5110. DOI: 10.3389/fncir.2012.00098.

[44] Gerard O'Leary et al. *OpenMEA: Open-Source Microelectrode Array Platform for Bioelectronic Interfacing*. en. Nov. 2022. DOI: 10.1101/2022.11.11.516234.

[45] Daejeong Kim, Hongki Kang, and Yoonkey Nam. "Compact 256-channel multi-well microelectrode array system for *in vitro* neuropharmacology test". en. In: *Lab on a Chip* 20.18 (2020), pp. 3410–3422. ISSN: 1473-0197, 1473-0189. DOI: 10.1039/D0LC00384K.

[46] Sagnik Middya et al. "Microelectrode Arrays for Simultaneous Electrophysiology and Advanced Optical Microscopy". en. In: *Advanced Science* 8.13 (July 2021), p. 2004434. ISSN: 2198-3844, 2198-3844. DOI: 10.1002/advs.202004434.

[47] Isaac A Weaver et al. "An open-source transparent microelectrode array". en. In: *Journal of Neural Engineering* 19.2 (Apr. 2022), p. 024001. ISSN: 1741-2560, 1741-2552. DOI: 10.1088/1741-2552/ac620d.

[48] Giovanni Pietro Seu et al. "Exploiting All Programmable SoCs in Neural Signal Analysis: A Closed-Loop Control for Large-Scale CMOS Multielectrode Arrays". In: *IEEE Transactions on Biomedical Circuits and Systems* 12.4 (Aug. 2018), pp. 839–850. ISSN: 1932-4545, 1940-9990. DOI: 10.1109/TBCAS.2018.2830659.

[49] Adam Paszke et al. *PyTorch: An Imperative Style, High-Performance Deep Learning Library*. en. Dec. 2019. DOI: 10.48550/arXiv.1912.01703.

[50] Jeffrey C. Magee. "Dendritic integration of excitatory synaptic input". en. In: *Nature Reviews Neuroscience* 1.3 (Dec. 2000), pp. 181–190. ISSN: 1471-003X, 1471-0048. DOI: 10.1038/35044552.

[51] Trym a. E. Lindell et al. "Information Encoding and Decoding in In Vitro Neural Networks on Micro Electrode Arrays through Stimulation Timing." en. In: *International Journal of Unconventional Computing* 20.3 (July 2025), p. 155. ISSN: 1548-7199. DOI: 10.32908/ijuc.v20.300924.